\documentstyle[floats,aps]{revtex}
\def\n{{\bf n}}
\begin{document}
\input epsf
\draft
\twocolumn[\hsize\textwidth\columnwidth\hsize\csname  
@twocolumnfalse\endcsname
\title{ Large N behavior of  Non-Ohmic Tunnel Junctions}

\author{S.R. Renn}
\address{
Department of Physics, University of California at San Diego,
La Jolla, CA 92093}

\date{\today}

\maketitle

\begin{abstract}
Non-Ohmic tunnel junctions are believed to occur in systems where the exciton
 and orthogonality catastrophe effects significantly modify tunneling rates. 
Here we present a simple non-perturbative treatment of the thermodynamic and
transport properties of subOhmic and superOhmic tunnel junctions. Our analysis
demonstrates the existence of a  quantum phase transition in 
subOhmic but not superOhmic tunnel junctions.
In addition, we find that the Coulomb gap vanishes 
continuously as the transition is approached.
\end{abstract}

\pacs{PACS numbers: 73.23.Hk, 71.10.Hf, 73.23.-b,73.40.Gk}
\vskip2pc]

\narrowtext

Recently there have been several  papers which have emphasized
 the importance of
the Fermi-edge singularities to electron tunneling phenomenon. These include 
a treatment of Coulomb blockade oscillations in double quantum dots by Matveev and Glazman\cite{Matveev} and
 the metal insulator transitions in granular materials 
by  Drewes, Renn and Guinea\cite{Drewes}.  In the later work, the 
authors argued  that the metal-insulator transition
  observed by Herzog et al\cite{Herzog} in
granular metallic wires was due to a quantum phase transition which
occurs in sub-Ohmic tunnel junctions.
Both these studies reflect the growing belief that  exciton
and orthogonality catastrophe effects are likely to be significant in 
a variety of systems including single tunnel junctions,  
 quantum dot arrays, and
  granular metallic composite materials.

The interest in  these  effects first began  with  the classic treatments 
of  x-ray absorption  by Mahan Noziers and
 De Dominicus(MND)\cite{mahan,Noziers}. However,
it was only a few years ago that the effect of Fermi edge singularities
on tunneling phenomenon was first considered. Perhaps the first work examine
these effects was by 
  Ueda and Guinea\cite{Guinea,Ueda} who  showed that  orthogonality catastrophe
 effects  could cause the conductance to behave as $G\sim T^x$ where $x>0$.
 A subsequent study of   resonant impurity tunneling by  Matveev and
 Larkin\cite{MatLark}
  predicted a zero temperature
 $I(V)$ characteristic  of the form
 $I(V) \propto (V-V_{th})^{-\alpha}\theta(V-V_{th})$. 
This prediction was  later confirmed\cite{Geim} in 
an experiment involving resonant tunneling between a 2-dimensional electron gas
 (2DEG) and a localized impurity.
Based on these works it is now  generally believed  that  
Fermi edge singularities a likely to produce a  variety of interesting 
 effects including  non-Ohmic
behavior in the $I(V)$ characteristics of tunnel junctions.
 For this reason, we will hereafter refer to
such systems as non-Ohmic tunnel junction (NOTJ's).

In order to better understand the behavior of NOTJ's,
we will present a  simple yet powerful non-perturbative treatment
 based on the large N expansion.
Our  approach provides 
convenient alternative to a renomalization group treatment of NOTJ's. In particular,
one can easily calculate  critical exponents characterizing the  metal Insulator transition which occurs in some subOhmic tunnel junctions. In addition, one can 
also obtain detailed expressions for conductances and specific heats. 
These calculations confirm the earlier prediction that sub-Ohmic tunnel junctions
exhibit a quantum phase  transition between an insulating  state to
a state with a  divergent zero temperature conductance. 
 In addition, the calculations indicate that the insulating state exhibits
a  renormalized Coulomb pseudo-gap
which collapses in a continuous fashion as the transition is approached.
Finally, the large-N approximation indicates that the metallic state is absent
in Ohmic tunnel junctions.

In order to  study
 the quantum dynamics of mesoscopic voltage fluctuations  in NOTJ,
will used a the long-range XY model introduced by Drewes et al.\cite{Drewes}
This  model is similar to that
first introduced by  Ben Jacob, Mottola,
 and Schon\cite{BenJacob} to treat Ohmic tunnel junctions.
The model is defined
by the  imaginary time partition function
\begin{equation}
Z=\int {\cal\ D}\n \ \delta(\n^2-1) \exp -S[\n]
\label{partition}\end{equation}
where $\n(\tau)=(\cos \phi(\tau), \sin (\phi(\tau))$ may be used to relate
 the XY spin orientation
to voltage using $\dot{\phi}(\tau)=eV/\hbar$.
As discussed by Drewes et al\cite{Drewes}, the effective action  of NOTJ
is
\begin{equation}
S[n]=\int d \tau d \tau' \alpha(\tau-\tau')[1-\n(\tau)\cdot \n(\tau')]
\end{equation}
where $\alpha(\tau)=\alpha_0\tau_Q^{-\epsilon}/\tau^{2-\epsilon}$.
The parameter  $\alpha_0$ is a dimensionless parameter 
proportional to the squared tunneling matrix element.
For $\epsilon=0$, this model reduces to that of Ben Jacob et al.

The above model has been analyzed Drewes et al 
using Monte Carlo simulations and the renormalization group.
The renormalization group analysis shows that the model exhibits a correlation time
\begin{equation}
\xi_{\tau} \sim |\alpha_0-\alpha_c|^{-1/\epsilon}
\end{equation}
where the phase transition occurs at $\alpha_c=1/2\pi^2\epsilon$. In the insulating phase one can associate $\hbar/\xi_{\tau}$ with a renormalized
 Coulomb gap subject to some important qualifications: (1.) The large $\alpha_0$ (spin-wave) calculation demonstrate $dI/dV \sim T^{2(1-\epsilon)}$ behavior. So we expect that a well defined Coulomb gap does not occur. However, at $T=0$ the leading order perturbation theory does give a threshold at $E_Q=e^2/2C$.
We expect therefore that higher order cotunneling processes will put states within the gap. So $\Delta \sim E_Q(\alpha_c-\alpha_0)^{-1/\epsilon}$ will not be a well defined gap although it may correspond to a pseudogap.

 The existence of pseudogap together with its behavior near the phase transition should  be observable  in the behavior of the   non-linear $I(V)$
 characteristics. More generally, non-linear $I(V)$ characteristics could
 provide a powerful means to test the scaling properties of the transition.
In particular,  one expects such  data according to the scaling ansatz
 scaling law\cite{Drewes} 
  \begin{equation}
dI/dV =\frac{e^2}{h}(k_BT/\Delta)^{\eta+1-\epsilon}F_{\pm}(eV/\Delta,k_BT/\Delta)
\end{equation}
 where $F_-(x,y)$ and $F_+(x,Y)$ are the $\alpha>\alpha_c$ and $\alpha<\alpha_c$ branches of  a universal scaling function which is finite for $x=y=0$. 
The $V\rightarrow 0$ limit of this scaling form has been verified using Monte 
Carlo finite size scaling methods by Drewes et al\cite{Drewesb}
al.

A final  consequence of the RG analysis  is the relation
 $\eta=1+\epsilon$\cite{Fisher}.
This result is very important since it  implies a temperature independent 
conductance at the subOhmic to insulator critical point. Although
 the conductance is  obtained from the 
fixed point Hamiltonian and is unaffected by irrelevant operators,
  it is not universal in any empirically useful sense.
This unfortunate result follows from the fact that the model exhibits a 
 line of fixed point Hamiltonians ${\cal H}_{\epsilon}$ which
enable the  ``universal conductances'' to take any value ranging  
from 0 to $\infty$. (See discussion below.)

At this point, we turn to our treatment of the  large N approximation. We
  begin with a generalization
of the long range XY model to a long range  Heisenberg model. The dynamical
degree of freedom of our  model is now an  $N$ component
 unit-vector spin-field, ${\hat n}=(n_1,n_2, \dots, n_N)$. 
Next we eliminate the $\hat{n}^2=1$ constraint by  introducing an auxiliary
 field $\lambda(\tau)$ in eq. \ref{partition}. This gives
\begin{eqnarray*}
Z=&\int {\cal D}[\n,\lambda] \ \exp -\int \frac{d \omega}{2 \pi} \frac{\alpha_0 C_N}{2}|\omega|^{1-\epsilon}|\n(\omega)|^2 \\
& -i\int d \tau  \lambda(\tau)[\n^2(\tau)-1]
\label{effective}\end{eqnarray*}
where $C_N=-4\tau_Q^{-\epsilon}\Gamma(\epsilon-1)\sin(\epsilon \pi/2)$.

As $N\rightarrow \infty$,  the functional integral is dominated by a saddle point on the imaginary $\lambda(\tau)$ axis\cite{comment}. Small fluctuations about the saddle point will correspond to a theory with a finite correlation length
$\xi_{\tau}=\hbar /\Delta$. Hence,  it will be convenient, to rewrite
the auxiliary field in the form
$\lambda(\tau)=\alpha_0C_N[u(\tau)- i \Delta^{1-\epsilon}/2]$
Next  we integrate out the $\n(\tau)$. This gives
\begin{eqnarray*}
S_{eff}(u)=&\frac{E_Q}{k_BT}(\frac{N}{2}-1) \ln(C_N\alpha_0)-\\
& \int_0^{\hbar \beta} d \tau {\alpha_0 C_N(\frac{\Delta^{1-\epsilon}}{2}
+i u(\tau})) \\
&+\frac{N}{2} \ln[|-i \partial_\tau|^{1-\epsilon}+\Delta^{1-\epsilon}+2i u(\tau)]
\end{eqnarray*}
 We then obtain $\Delta$ by solving  $\delta S_{eff}[u]/\delta u(\tau)|_{u=0}=0$ or equivalently by solving
\begin{equation}
\alpha_0C_N =N \int \frac{d \omega}{2 \pi} \frac{1}{|\omega|^{1-\epsilon} +\Delta^{1-\epsilon}}\approx\frac{N}{\epsilon \pi}\left[E^{\epsilon}_Q-\Delta^{\epsilon}\right]
\end{equation}
 This gives the 
\begin{equation}
\Delta=E_Q (1 -\frac{\alpha_0}{\alpha_c})^{1/\epsilon}
\label{delteq}\end{equation}
where
\begin{equation}
\alpha_c=-\frac{N}{4\pi \epsilon}\frac{1}{\Gamma(\epsilon-1)\sin(
\frac{\epsilon \pi}{2})}
\end{equation}
The $\epsilon \rightarrow 0$ limit, behavior of the $\Delta$ is
\begin{equation}
\Delta=E_Q\exp -\frac{2\pi^2 \alpha_0}{N}
\end{equation}

Next we consider spin-fluctuations about the saddle point. 
In the large N  limit, $G(\tau-\tau')=<\n(\tau)\cdot \n(\tau')>$ is given by
\begin{equation}
G(\omega)=\frac{N}{\alpha_0 C_N}\frac{1}{|\omega|^{1-\epsilon}+ \Delta^{1-\epsilon}}
\end{equation}
This result demonstrates that $\xi_{\tau}=\hbar/\Delta$ is indeed a correlation length.  The fact that $\xi_{\tau}$ diverges when $\alpha_0=\alpha_c$ then implies that $\alpha_c=N/2\pi^2\epsilon$ is a critical point. This 
leading order in $1/N$ result is, of course, consistent with the
 renormalization group results that $\alpha_c=(N-1)/2\pi^2 \epsilon [1+O(\epsilon)]$. At the critical point the spin-spin correlation function becomes
\begin{equation}
G(\tau)=\Gamma(\epsilon+1)\cos(\frac{\epsilon \pi}{2}) 
\left(\frac{\tau_Q}{\tau}\right)^{\epsilon}
\label{critG}\end{equation}
This form  implies $\eta=1+\epsilon$ as was previously
  obtained by Fisher Ma and Nickle\cite{Fisher} and is believed to be exact.

Next we wish to consider the behavior of the dimensionless XY model specific heat. The 
quantity under discussion is
 $C_{XY}\equiv \alpha_0^2 d^2 \ln Z/d \alpha_0^2$. This is not the physical specific heat of a tunnel junction. It is, however, a
quantity which is relevant to Quantum Monte Carlo studies\cite{Scalia} of the
 tunnel junction. 
The calculation the heat capacity proceeds as follows. First, we calculate
$F\equiv -k_BT\ln Z$ to leading order in $1/N$ using the $u(\tau)=0$ saddle point.
This gives a zero temperature free energy
\begin{eqnarray*}
F=&\frac{N}{2}E_Q\ln (C_N \alpha_0) -\frac{1}{2} \alpha_0 C_N \Delta^{1-\epsilon}\\
& + \frac{N}{2} \int \frac{d \omega}{2 \pi} \ln \left(|\omega|^{1-\epsilon}+\Delta^{1-\epsilon} \right)
\end{eqnarray*}
This coincides with $k_BTS[u=0]$.
Now using the Euler-Lagrange equation for $\Delta$ together with eq. \ref{delteq} one obtains the expression
\begin{equation}
\frac{C_{XY}}{N \beta E_Q} =\frac{1}{2} - \frac{1}{2\pi}\frac{1-\epsilon}{\epsilon^2}
\left(\frac{\alpha_0}{\alpha_c}\right)^2 \left(1-\frac{\alpha_0}{\alpha_c}\right)^{\frac{1}{\epsilon}-2}
\end{equation}
This result is valid to leading order in $O(1/N)$ provided that  $\alpha_0<\alpha_c$ and $|\alpha_c-\alpha_0|/\alpha_0$ are both small. One sees that the specific heat exponent $\alpha=2-1/\epsilon$ coincides with that  
obtained from the Josephson  relation $\alpha=2-d\nu$.

At this point, we wish to consider the behavior of the dc conductance. First we generalize the electric current.  For the physical $U(1)=O(2)$ model, the current is
 obtained using $I(\tau)=-\delta S[A_x]/\delta A_x(\tau)$ where
\begin{equation}
\frac{S[A_x]}{\hbar}= \int d\tau d\tau' \alpha(\tau-\tau')[1- 
 \cos(\Phi(\tau)-\Phi(\tau'))]
\end{equation}
where $\Phi(\tau)\equiv \phi(\tau)-\frac{e}{\hbar}A_x(\tau)$.  This gives
\begin{equation}
I(\tau)=2e\int d\tau' \alpha(\tau-\tau')\sin(\phi(\tau)-\phi(\tau'))
\label{current}\end{equation}
Equivalently, an expression for the  electric current could be obtained
  by considering the variation of $S$ under a local gauge transformation of the form 
$\delta \phi(\tau) = \delta \epsilon(\tau)$.
Then $\frac{e}{\hbar}\delta S/\delta \epsilon(\tau)=I(\tau)$. This second definition is convenient for   defining the electric current(s) in the $O(N)$
model.
In the O(N) model one has a set of  $N-1$ electric currents, $I^a(\tau)$, $a=1 \dots N-1$ each associated with the $N-1$ Lie algebra generators ${\bf T}^a$.  Since ${\bf n}(\tau)$ is an  $N$ component vector field which transform according to the fundamental representation of $O(N)$, the  ${\bf T}^a$, matrices are  $N\times N$ anti-symmetric matrices which can be chosen such that ${\rm Tr}  {\bf T}^a{\bf T}^b=\delta_{ab}$.
 We now define the electric currents as follows:
 Under a local gauge transformation $\delta {\bf n}(\tau)/\delta \epsilon^a (\tau')={\bf T}^a\cdot {\bf n}(\tau)\delta (\tau-\tau')$. The currents are then defined to
 be $I^a(\tau)=\frac{e}{\hbar}\delta S[\epsilon]/\delta \epsilon^a(\tau)$. From this we obtain the explicit 
form of the electric currents
\begin{equation}
I^a(\tau)= 2e \int d\tau' \alpha(\tau-\tau') \vec{n}(\tau)\cdot {\bf T}^a \cdot \vec{n}(\tau')
\end{equation}
The special case of the $O(2)=U(1)$ model, the above result reduces to eq. \ref{current} as 
required. This may be shown by taking ${\bf T}=i\sigma_y$ and ${\bf n}(\tau)=(\cos(\phi(\tau)),\sin(\phi(\tau')))$.

Having defined the $O(N)$ electric currents, we want to perform a Kubo calculation
of the dc conductances $G^{ab}$. The first step is to use the identity

\begin{equation}
<\frac{\delta S}{\delta \epsilon^a(\tau)}\frac{\delta S}{\delta \epsilon^b(\tau') }>=-<\frac{\delta^2 S}{\delta \epsilon^a(\tau)\delta \epsilon^b(\tau')}>
\end{equation}
which follows from a functional integration by parts.
This gives a current current correlation function for the physical model (i.e. $U(1)=O(2)$ model) of the form\cite{Simanek}
\begin{equation}
<I^a(\tau)I^b(\tau')>=-\frac{2e^2}{\hbar}\alpha(\tau-\tau')<\n(\tau)\cdot {\bf T}^a {\bf T}^b \cdot \n(\tau')>
\label{jj}
\end{equation}

Using the Kubo formula and eqs. \ref{critG} and \ref{jj} we  obtain the
 critical conductance, $G_c$, of the tunnel junction
 to leading order in $1/N$:
\begin{equation}
G^{ab}=2\pi (1-\epsilon){\rm ctn} \left(\frac{\epsilon \pi}{2}\right)\frac{e^2}{h}\delta_{ab}=G_c\delta_{ab}
\label{Gc}\end{equation}
Because of the value of the $\eta$ exponent the conductance is temperature independent. 
Next we observe  that the conductance diverges as $\epsilon \rightarrow 0$. See
fig. 1.
This means that for small $\epsilon$ a rather large value of   $\alpha_0$ (or equivalently the tunneling matrix element) is required in order to obtain a subohmic phase. This is consistent with the belief that the ordered (subOhmic) phase is destroyed as $\epsilon \rightarrow 0$ and that the phase transition is absent for positive $\epsilon$.  The second observation is that as $\epsilon \rightarrow 1$ $\sigma 
\rightarrow 0$.  This is consistent with the fact that the disordered (insulating) phase is absent for $\epsilon \ge 1$. In particular, it implies that an arbitrarily weak $\alpha_0$ will order the XY model as $\epsilon \rightarrow 1$ is approached. 
The fact that the critical conductance can, depending on the value of $\epsilon$, range from $0$ to $\infty$  is interesting since it implies that 
 no universal value of the conductance should be expected.
\begin{figure}[h]
\centering
\leavevmode
\epsfxsize=9cm
\epsfbox[18 18 552 482] {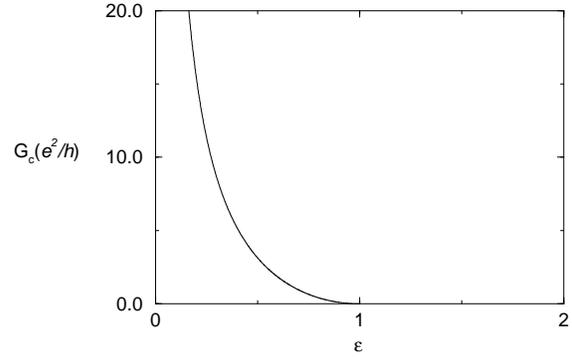}
\caption[]
{ The critical conductance vs. $\epsilon$. Observe that the
insulating phase is squeezed out of the phase diagram (i.e. $G_c\rightarrow 0$) as $\epsilon\rightarrow 1$. Similarly the conducting phase is squeezed out  as $\epsilon \rightarrow 0$.
}
\end{figure}

Next we consider the  AC conductance $G(\omega)$. For this purpose it is convenient to approximate $<{\bf n}(\tau)\cdot {\bf n}(0)>$ as being given by its critical behavior for $\tau<<1/\Delta$ and exhibiting a cutoff for $\tau \sim 1/\Delta$ i.e. we assume that
\begin{equation}
G(\tau) \approx \Gamma(1+\epsilon)\cos(\frac{\epsilon \pi}{2})
\frac{\tau_Q^{\epsilon}}{\tau^{\epsilon}} \exp-(\Delta |\tau|)
\end{equation}
With this approximate form one can then obtain the time-ordered real time current current correlation function:
\begin{equation}
C_t^{ab}(\omega)\equiv -i \int^{\infty}_{-\infty}dt \ <T_t(I^a(\tau)I^b(0))> \exp {i\omega t}
\end{equation}
Using eq. \ref{jj} one obtains
\begin{equation}
C_t^{ab}(\omega) =-2\Delta G_c \delta_{ab} \int^{\infty}_{-\infty} \frac{d\omega'}{2\pi} \frac{|\omega'|}{(\omega'-\omega)^2-\Delta^2 +\i \delta}
\label{Ct}\end{equation}
 This   in conjunction with the  the Kubo formula is then used to obtain the 
 the ac conductance $G(\omega)^{ab}=[G'(\omega)+i G''(\omega)]\delta_{ab}$. The real part is
 found to be
\begin{equation}
G'(\omega)=G_c(1-\frac{\Delta}{2|\omega|})\theta(|\omega|-\Delta)
\label{Greal}
\end{equation}
where $\theta(x)=0$ for $x<0$ and 1 for $x>0$ and where $G_c$ is the critical conductance as given by eq. \ref{Gc}.  This result allows one to 
identify $\Delta$ as an excitation gap which, near the transition, is much smaller than the charging energy $E_Q$.

Previously, Drewes et al\cite{Drewes} found a well defined Coulomb gap in 
 a leading order ( $O(\alpha_0)$)  perturbative calculation of the 
 $I(V)$ characteristic. However a $\alpha_0^2$ calculation of $G(T)$
found that
\begin{equation}
G\sim \frac{\alpha_0^2}{R_Q}\left(\frac{\pi k_BT}{E_Q}\right)^{2(1-\epsilon)}
\end{equation}
 That  result indicates that the conductance does not exhibit an 
 activated  temperature dependence and that 
 higher order processes introduce states inside the gap. Moreover, one would argue
that the  zero temperature
 $I(V)$ characteristic is of the form 
 $I(V)\sim (e^2/h)\alpha_0^2 (eV/\Delta)^{2(1-\epsilon)}$
for $V<<\Delta$ and that ac conductance is of the form 
 \begin{equation}
G'(\omega)\sim (\omega/\Delta)^{2(1-\epsilon)}
\end{equation}
when $\omega<<\Delta$.
These results   presumable occur at  higher than leading
 order  in the $1/N$ expansion.

In addition to calculating the $G'(\omega)$ one can also calculate the
imaginary  part of the conductance.
In this case one finds
\begin{equation}
G''(\omega)=\frac{C_t(0)}{\omega} +\frac{\Delta}{\pi \omega}G_c\left[ \ln\left|
\frac{\Delta^2}{\Delta^2-\omega^2}\right|+
\frac{\omega}{\Delta}\ln\left|\frac{\omega-\Delta}{\omega+\Delta}\right|
 \right]
\end{equation}
where $C_t(0)$ depends logarithmically on $E_Q/\Delta$.

\noindent {Acknowledgements} The author would like to acknowledge support from
 NSF Grant No. DMR 91-13631,  the Hellman foundation,
 the Alfred P. Sloan Foundation . We would like to acknowledge
useful conversations with D. Arovas, F. Guinea, and  S. Drewes.


\begin{references}
\bibitem{Matveev}K.A. Matveev, L.I. Glazman, H.U. Baranger, preprint cond-mat 9504099.
\bibitem{Drewes}S. Drewes, S. Renn, and F. Guinea, UCSD preprint (1997).

\bibitem{Herzog} A. V. Herzog, P. Xiong, F. Sharifi, and R.C. Dynes, Phys. Rev Lett. {\bf 76}, 668 (1996).

\bibitem{mahan} G.D. Mahan, Phys. Rev. {\bf 153}, 882 (1967); G.D. Mahan, Phys. Rev. {\bf 163}, 612 (1967).
\bibitem{Noziers}P. Nozi\'eres and C. T. De Dominicis, Phys. Rev. {\bf 178}, 1097 (1969).
\bibitem{Guinea}M. Ueda and F. Guinea, Z. Phys. B {\bf 85}, 413 (1991).
\bibitem{Ueda} M. Ueda and S. Kurihara in Macroscopic quantum phenomena, T.D. Clark, H. Prance, R.J. Prance, T.P. Spiller (eds.), p.143, Singapore, World Scientific (1990).
\bibitem{MatLark} K.A. Matveev, A.I. Larkin, Phys. Rev. B {\bf 46}, 15337 (1992).
\bibitem{Geim}A.K. Geim, P.C. Main, N. La Scala, Jr., L. Eaves, T.J. Foster, P.H. Beton, J. W. Sakai, F. W. Sheard, M. Henini, G. Hill, M.A. Pate , Phys. Rev. Lett. {\bf 72}, 2061 (1994).
\bibitem{BenJacob}E. Ben-Jacob, E. Mottola, and G. Sch$\ddot{o}$n, Phys. Rev. Lett. {\bf 51}, 2064 (1983).

\bibitem{Drewesb}S. Drewes and S. Renn, work in progress.
\bibitem{Fisher}M.E. Fisher, Shang-keng Ma, B.G. Nickel, Phys. Rev. Lett. {\bf 29}, 917 (1972).
\bibitem{comment} This follows since $\delta^2 S_{eff}[u]/\delta u(\tau) \delta u(\tau') \propto N$, where $S_{eff}[u]$ 
and  $u(\tau)$ are defined below.
\bibitem{Scalia}V. Scalia, G. Falci, R. Fazio, G. Giaquinta, Z. Phys. B. {\bf 85}, 427-433(1991).

\bibitem{Simanek} R. Brown and E. Simanek, Phys. REv. B {\bf 34}, 2957 (1986);


\end{references}
\end{document}